\documentclass[10pt,numbers,preprint,nocopyrightspace]{sigplan/sigplanconf}
\usepackage{amsmath}

\usepackage[utf8x]{inputenc}
\usepackage[T1]{fontenc}
\usepackage{microtype}

\usepackage[inline]{enumitem}

\usepackage{flushend}

\usepackage{placeins}

\usepackage{siunitx}
\DeclareSIUnit[number-unit-product = \,] \factor{x}
\usepackage[smaller]{acronym}
\newcommand{\HAMT}{\ac{HAMT}\xspace}
\acrodef{HAMT}{Hash-Array Mapped Trie}

\acrodef{AMT}{Array Mapped Trie}

\newcommand{\CHAMP}{\ac{CHAMP}\xspace}
\acrodef{CHAMP}{Compressed Hash-Array Mapped Prefix-tree}

\newcommand{\HCHAMP}{\ac{HCHAMP}\xspace}
\acrodef{HCHAMP}{Heterogeneous Compressed Hash-Array Mapped Prefix-tree}

\newcommand{\HHAMT}{\ac{HHAMT}\xspace}
\acrodef{HHAMT}{Heterogeneous Hash-Array Mapped Trie}

\acrodef{MEMCHAMP}{MEMoized Compressed Hash-Array Mapped Prefix-tree}

\newcommand{\JIT}{\ac{JIT}\xspace}
\acrodef{JIT}{Just-in-time}

\newcommand{\JDK}{\ac{JDK}\xspace}
\acrodef{JDK}{Java Development Kit}

\newcommand{\JVM}{\ac{JVM}\xspace}
\acrodef{JVM}{Java Virtual Machine}

\newcommand{\VM}{\ac{VM}\xspace}
\acrodef{VM}{Virtual Machine}

\acrodef{JVMTI}{Java Virtual Machine Tool Interface}

\newcommand{\GC}{\ac{GC}\xspace}
\acrodef{GC}{Garbage Collector}

\acrodef{BCI}{Bytecode Instrumentation}

\acrodef{PDB}{Program Data Base}

\acrodef{DSL}{Domain-Specific Language}

\acrodef{SLOC}{Source Lines of Code}

\acrodef{DAG}{Directed Acyclic Graph}

\acrodef{REPL}{Read–Eval–Print Loop}

\acrodef{AOP}{Aspect-Oriented Programming}

\newcommand{\API}{\ac{API}\xspace}
\acrodef{API}{Application Program Interface}

\acrodef{ADT}{Algebraic Data Type}

\newcommand{\JMH}{\ac{JMH}\xspace}
\acrodef{JMH}{Java Microbenchmarking Harness}

\newcommand{\CFG}{\ac{CFG}\xspace}
\acrodef{CFG}{Control-Flow Graph}

\newcommand{\LLC}{\ac{LLC}\xspace}
\acrodef{LLC}{Last-Level Cache}

\acrodef{STM}{Software Transactional Memory}

\acrodef{AST}{Abstract Syntax Tree}

\acrodef{DFA}{Deterministic Finite Automaton}

\acrodef{CMS}{Content Management System}

\newcommand{\MAD}{\ac{MAD}\xspace}
\acrodef{MAD}{Median Absolute Deviation}

\acrodef{ORP}{Object Redundancy Profiling}

\acrodef{ECP}{Equals-Call Profiling}

\acrodef{MSM}{Maximal Sharing Model}

\acrodef{OEP}{Object Equality Profiling}

\newcommand{\FIXNUM}{\ac{FIXNUM}\xspace}
\acrodef{FIXNUM}{Fixed-Width Numeric}

\newcommand{\BIGNUM}{\ac{BIGNUM}\xspace}
\acrodef{BIGNUM}{Big Object-Represented Numeric}

\newcommand{\CPU}{\ac{CPU}\xspace}
\acrodef{CPU}{Central Processing Unit}

\newcommand{\CSV}{\ac{CSV}\xspace}
\acrodef{CSV}{Comma-Separated Values} % Plural Acronym!

\usepackage{fixltx2e}

\usepackage{xspace}

\usepackage{tikz}
\usetikzlibrary{arrows,calc,patterns,positioning,shapes}

\usepackage[caption=false]{subfig}
\usepackage{rotating}
\usepackage{rotfloat}
\usepackage{pdflscape}
\usepackage{tabularx}
\newcolumntype{R}{>{\raggedleft\arraybackslash}X}%
\newcolumntype{C}{>{\centering\arraybackslash}X}%

\newcommand{\tableheader}[1]{\textbf{#1}}%
\usepackage{booktabs}
\newcommand{\ra}[1]{\renewcommand{\arraystretch}{#1}}

\usepackage{multirow}

\usepackage{setspace}

\usepackage[scaled=0.82]{beramono}
\usepackage{color}
\definecolor{bluekeywords}{rgb}{0.13,0.13,1}
\definecolor{greencomments}{rgb}{0,0.5,0}
\definecolor{redstrings}{rgb}{0.9,0,0}
\definecolor{gray}{rgb}{0.5,0.5,0.5}
\definecolor{ggray}{gray}{0.2}

\definecolor{Gray}{gray}{0.9}
\definecolor{LightGray}{gray}{0.4}
\definecolor{VeryLightGray}{gray}{0.8}

\usepackage{listings}

\usepackage{calc}

\makeatletter
\newlength{\linenumwidth} 
\newlength{\numwidth}%
\def\lst@PlaceNumber{%
  \makebox[\linenumwidth][l]{%
    \makebox[\numwidth][r]{\normalfont\tiny\lst@numberstyle{\thelstnumber}}%
  }%
}
\makeatother

\newcount\bsubfloatcount
\newtoks\bsubfloattoks
\newdimen\bsubfloatht

\usepackage{layouts}
\usepackage{hyperref}

\begin{document}

\setlength{\pdfpageheight}{\paperheight}
\setlength{\pdfpagewidth}{\paperwidth}

\toappear{}

%\conferenceinfo{OOPSLA '15}{October 25--30, 2015, Pittsburgh, Pennsylvania, USA} 
%\copyrightyear{20yy} 
%\copyrightdata{978-1-nnnn-nnnn-n/yy/mm} 
%\doi{nnnnnnn.nnnnnnn}

% Uncomment one of the following two, if you are not going for the 
% traditional copyright transfer agreement.

%\exclusivelicense                % ACM gets exclusive license to publish, 
                                  % you retain copyright

%\permissiontopublish             % ACM gets nonexclusive license to publish
                                  % (paid open-access papers, 
                                  % short abstracts)

\titlebanner{DRAFT}        % These are ignored unless
\preprintfooter{Fast and Lean Immutable Multi-Maps}   	% 'preprint' option specified.

\title{Fast and Lean Immutable Multi-Maps on the JVM based on Heterogeneous Hash-Array Mapped Tries}
%\subtitle{Optimizing Heterogeneous Hash-Array Mapped Tries}
%\subtitle{A Memory-Efficient Encoding for (Immutable) Collections of Either-Types}
%\titlerunning{Optimizing Hash Array Mapped Tries}

%%% Alternative titles:
% Heterogeneous Data Structures for Optimizing Dynamic Language Runtimes
% Array-less is More
% Self-Adaptive Heterogeneous Data Structures for Optimizing Dynamic Language Runtimes

\authorinfo{Michael J. Steindorfer}
           {\href{http://www.cwi.nl/sen1}{Centrum Wiskunde \& Informatica, The Netherlands}}
           {\href{mailto:Michael.Steindorfer@cwi.nl}{Michael.Steindorfer@cwi.nl}}
\authorinfo{Jurgen J. Vinju}
           {\href{http://www.cwi.nl/sen1}{Centrum Wiskunde \& Informatica, The Netherlands} \\
           TU Eindhoven, The Netherlands \\
           INRIA Lille, France}
           {\href{mailto:Jurgen.Vinju@cwi.nl}{Jurgen.Vinju@cwi.nl}\vspace{2.8mm}} % \vspace{-5mm}

%\authorinfo{Name1}
%           {Affiliation}
%           {Email}
%\authorinfo{Name2}
%           {Affiliation}
%           {Email}

\maketitle

\begin{abstract}
% !TEX encoding = UTF-8 Unicode
% !TEX root = ../paper.tex

An immutable multi-map is a many-to-many thread-friendly map data structure with expected fast insert and lookup operations. This data structure is used for applications processing graphs or many-to-many relations as applied in static analysis of object-oriented systems. When processing such big data sets the memory overhead of the data structure encoding itself is a memory usage bottleneck. Motivated by reuse and type-safety, libraries for Java, Scala and Clojure typically implement immutable multi-maps by nesting sets as the values with the keys of a trie map. Like this, based on our measurements the expected byte overhead for a sparse multi-map per stored entry adds up to around \SI{65}{\byte}, which renders it unfeasible to compute with effectively on the \acs{JVM}.

In this paper we propose a general framework for \acused{HAMT}\aclp{HAMT} on the \acs{JVM} which can store type-heterogeneous keys and values: a \HHAMT. Among other applications, this allows for a highly efficient multi-map encoding by \begin{enumerate*}[label=(\alph*)]\item not reserving space for empty value sets and \item inlining the values of singleton sets while maintaining a \item type-safe \acs{API}\end{enumerate*}. 

We detail the necessary encoding and optimizations to mitigate the overhead of storing and retrieving heterogeneous data in a hash-trie. Furthermore, we evaluate \HHAMT specifically for the application to multi-maps, comparing them to state-of-the-art encodings of multi-maps in Java, Scala and Clojure. We isolate key differences using microbenchmarks and validate the resulting conclusions on a real world case in static analysis. The new encoding brings the per key-value storage overhead down to \SI{30}{\byte}: a \SI{2}{\factor} improvement. With additional inlining of primitive values it reaches a \SI{4}{\factor} improvement.
\end{abstract}

% \category{E.2}{DATA STORAGE REPRESENTATIONS}{Hash-table representations}

% general terms are not compulsory anymore, 
% you may leave them out
%\terms
%term1, term2

% \keywords{Data structures, heterogeneity, memory optimization, performance, immutability, persistent data structures, specialization, code generation, \acl{JVM}.}

\acresetall % reset all acronyms
%\input{graphics/tikz_prelude.tex}
% !TEX encoding = UTF-8 Unicode
% !TEX root = ../paper.tex

\section{Introduction}

This paper is about the challenges of optimizing immutable multi-maps on the \JVM and how they can be solved using a general method of coding heterogenous hash-array mapped tries. A multi-map is a data structure which acts as an associative array storing possibly multiple values with a specific key. Typically multi-maps are used to store graphs or many-to-many relations.

Many-to-many relations or graphs in general occur naturally in application areas such as static analysis of object-oriented software. In some applications it is the case that the initial raw data is many-to-one, and further processing or exploration incrementally leads to a many-to-many mapping for some of the entries. In other applications the distribution of sizes of the range sets in the raw data is highly skewed, such as when representing scale-free networks, like academic citations, the web, online social networks, and program dependence graphs. The number of values associated with a specific key is then practically always very low, yet there are possibly numerous exceptions to cater for nevertheless, where many values end up being associated with the same key. A key insight in the current paper is that we can exploit these highly common skewed distributions to save memory for the most frequent cases.

On the \JVM relations are not natively language-supported; rather the standard libraries of Java, Scala and Clojure either provide implementations of multi-maps, or the map and set \acp{API} allow programmers to construct multi-maps easily in a type-safe manner (i.e., using sets as the values of a normal polymorphic map). The goal of this paper is to overcome the limitations of these existing implementations of multi-maps, improving drastically on the memory footprint without loss of storage, lookup and iteration efficiency. Typically state-of-the-art multi-maps come with a mode of \SI{65}{\byte} overhead per stored key/value item, while the most compressed new encoding in this paper reaches an optimum of \SI{30}{\byte}. In general the encoding has \SI{2}{\factor} smaller footprints (modal) when storing reference objects, and \SI{4}{\factor} smaller footprints when storing Java primitive values.

On the \JVM, immutable collections are used mostly by functional/object-oriented programmers from the Scala and Clojure communities. However, since Java~8 the functional and streaming \acp{API}~\cite{DBLP:conf/ecoop/BiboudisPFS15} are becoming mature, making immutable collections become more relevant in the Java context. Immutability for collections implies referential transparency (without giving up on sharing data) and it satisfies safety requirements for having co-variant sub-types~\cite{Igarashi:2002hk}. Because of these properties, immutable collections are also safely shared in presence of concurrency. 

Our point of departure is the \HAMT data structure~\cite{Bagwell:2001tw}, which has proven to be an efficient immutable alternative to array-based implementations. In contrast to arrays, \acp{HAMT} enable fine-grained memory layout optimizations~\cite{Steindorfer:2014dk}. There exists an optimized encoding~\cite{Steindorfer:2015iz} of \acp{HAMT} tailored the \JVM, named \CHAMP. The \CHAMP data structure allows for time and memory efficient immutable maps and sets. To efficiently encode multi-maps we propose a generalisation of the \CHAMP data structure to allow for heterogeneous data shapes. The new resulting data structure, called \HHAMT, unifies design elements from both \HAMT and \CHAMP. A \HHAMT allows for a type-safe \API in which keys and values can be represented using different types of data within the same map. This allows for all kinds of optimized data structures, but we focus on multi-maps in this paper as the key purpose. A basic dichotomous \HHAMT multi-map is used to either store an inlined single value, or a full nested set data structure. We propose an efficient encoding of \HHAMT to mitigate the incurred overhead. 

\subsection{Contributions and Roadmap}
We address the design and evaluation of \HHAMT as follows:
\begin{itemize}
\item Section~\ref{sec:conribution_heterogeneous_data_layout} describes the foundations of \HHAMT and identifies the main sources of overhead that need to be mitigated. 
\item Section~\ref{sec:conribution_scalable_specialization_encoding} outlines scalable encoding of source code specializations (and their necessary runtime support) to yield memory savings between \SI{2}{\factor} and \SI{4}{\factor}.
\item Section~\ref{sec:evaluation-internal} compares \HHAMT against \CHAMP (baseline) to understand the cost of turning a (homogeneous) map into a (heterogeneous) multi-map.
\item Section~\ref{sec:evaluation-external} compares a specialized \HHAMT multi-map against idiomatic solutions from Clojure and Scala.

\item Section~\ref{sec:evaluation-primitive-collections} compares the memory footprint of a specialized \HHAMT multi-map against state-of-the-art primitive collection libraries (Goldman Sachs, FastUtil, Trove,  Mahout). 
\item Section~\ref{sec:evaluation-real-world} compares the performance of multi-maps in \HHAMT, Clojure, and Scala on a realistic case.
\end{itemize}
Section~\ref{sec:related_work} discusses related work and Section~\ref{sec:further-applications} enumerates further use cases for heterogeneity, before we conclude in Section~\ref{sec:conclusion}. 
%\input{chapters/background.tex}
% !TEX encoding = UTF-8 Unicode
% !TEX root = ../paper.tex

\section{Heterogeneous Hash-Trie Data Layout\label{sec:conribution_heterogeneous_data_layout}}
\begin{figure}[t]
	\vspace{-2mm}
	\centering
	\begin{lstlisting}[escapechar=!, label=lst:skeleton, caption=Skeletons of a various \acp{HAMT}.]
abstract class HamtCollection {!\label{HAMTstart}!
	HamtNode root; int size;
	// 1-bit + runtime checks (e.g., instanceof)
	class HamtNode {
	  int bitmap; 
	  Object[] contentArray;
	}
}!\label{HAMTend}!
abstract class ChampCollection {!\label{CHAMPstart}!
	ChampNode root; int size;

	// 2-bits (distributed)
	class ChampNode {
	  int datamap; 
	  int nodemap;
	  Object[] contentArray;
	}
}!\label{CHAMPend}!
abstract class HeterogeneousHamtCollection {  !\label{HHAMTstart}!
	 HeterogeneousHamtNode root; int size;

	// n-bits (consecutive)
	class HeterogeneousHamtNode {
	  BitVector bitmap = new BitVector(n * 32);
	  Object[] contentArray;
	}
}!\label{HHAMTend}!
	\end{lstlisting}
	\vspace{-5mm}
\end{figure}

A general trie~\cite{DeLaBriandais:1959kp,Fredkin:1960eq} is a lookup structure for finite strings that acts like a Deterministic Finite Automaton (DFA) without any loops: the transitions are the characters of the strings, the internal nodes encode prefix sharing, and the accept nodes may point to values associated with the strings. In a \ac{HAMT}, the strings are the bits of the hash codes of the elements stored in the trie. A \HAMT is memory efficient not only because prefixes are shared, but also because child nodes are only allocated if the prefixes of two or more elements overlap. 

The first class in Listing~\ref{lst:skeleton}~(lines~\ref{HAMTstart}--\ref{HAMTend}) depicts a typical encoding of a \HAMT in Java. A single \num{32}-bit integer bitmap is used to encode which of the 32 trie-branches ---and correspondingly which slots in the untyped array--- are used, together with a mapping function that calculates offsets in the array by counting bits in the bitmap. In general, a \HAMT must be able to distinguish between three possible states for each trie-branch: absence of data, and otherwise distinguishing the data category (either payload, or a sub-node). Because a single bit cannot differentiate three different states, additional dynamic checks ---such as \lstinline{instanceof}--- are used for discriminating the data category. Note that data payload and sub-nodes occur in arbitrary order in the array. 

The second class in Listing~\ref{lst:skeleton}~(lines~\ref{CHAMPstart}--\ref{CHAMPend}) depicts the skeleton of the \CHAMP encoding~\cite{Steindorfer:2015iz}, which operates like a \HAMT but uses an explicit encoding to eliminate dynamic \lstinline{instanceof} checks. With two bitmaps \CHAMP improves the mapping function to regroup the array slots into two separate homogeneously-typed sequences: a sequence of data payload, followed by a sequence of sub-node references. Because each homogeneous sequence uses its own bitmap, \CHAMP kept the bitmap processing identical to \acp{HAMT}.

\paragraph{Summary.} In a \HAMT, each trie node contains an arbitrary mix of data elements and sub-nodes, therefore array slots require type checks individually. In contrast, \CHAMP splits \HAMT's mixed data sequence into two homogeneous sequences, enabling optimizations that were not possible before. A key to performance ---when iterating over or batch-processing elements of homogeneous or heterogeneous data structures--- is that individual elements do not need to be checked for its specific type~\cite{Bolz:2013ku}. This is also one of the reasons why the \CHAMP performs better than the \HAMT.
In short: the homogeneous \CHAMP data structure provides a good starting point for heterogeneous collections.

\subsection{Generalizing Towards a Heterogeneous \HAMT}

The third class in Listing~\ref{lst:skeleton}~(lines \ref{HHAMTstart}--\ref{HHAMTend}) illustrates the proposed \HHAMT skeleton. \HHAMT uses a multi-bit encoding like \CHAMP but reverts to a sequential representation: one larger bitmap that stores a sequence of \num{32} $n$-bit tuples consecutively, instead of at maximum $k$ individual bitmaps. $k$ denotes the maximum number of supported heterogeneous types while $n$ denotes the number of bits needed in our encoding.

For any $k$, a \HHAMT requires $n = \lceil{}log_2(k + 2)\rceil$ bits at minimum per trie-branch to encode all of its possible states. The two additional states are needed for encoding the absence of a trie branch, and encoding sub-trees in case of hash-prefix collisions.
For the sake of clarity we mainly focus on the $k=2$ case in the evaluation (Sections~\ref{sec:evaluation-internal}, \ref{sec:evaluation-external}, \ref{sec:evaluation-primitive-collections} and \ref{sec:evaluation-real-world}), where the required number of bits $n = 2$. This case covers the scenario of distinguishing between a singleton value, and an arbitrarily sized nested set for multi-map implementations. However in the current section we detail the general design and code for arbitrary $k$. Note that fixing $k$ does influence efficiency trade-offs: experimental results for $k = 2$ do not generalize directly to other values of $k$.

\subsection{\HHAMT \API}

Although this is not a core contribution, since we model data structures beyond the power of Java's type system, we should detail how to circumvent it. Java does not support union types, and a polymorphic wrapper (such as Scala's \lstinline{Either}) would introduce overhead. To solve this  we can either write or generate specialized code for fixed combinations of types, or use Java's method type \emph{polymorphism} and judicious use of class literals (a.k.a. type tokens like \lstinline{Integer.class}). 

For multi-maps, which are heterogeneous only internally, a simple generic \API will suffice.
For other applications, such as when the keys or values of a map are type heterogeneous or primitive values are inlined, code generation for the wrapping \API is possible. 
If we use Java's method polymorpism~(cf. \emph{Effective Java}, Item 29 \cite{effective-java}) instead we may avoid code generation at a certain cost. We use type tokens and their \lstinline{cast} method to encode type heterogeneity. Up to Java 8 it is not possible to bind primitive types to type variables though, and care must be taken to avoid dynamic type errors. Casts can be avoided using either-typed (temporary) wrappers or a typed callback interface. Examples can be found in the Appendix. Note that the internals of the \HHAMT can always decide upon the type a value with \SI{100}{\percent} certainty. % Appendix~\ref{sec:generic-api}

\subsection{Bitmap Encoding and Indexing}
\newcommand{\etcetera}{\emph{\ldots code for lookup, insert or delete \ldots}}

\begin{figure*}[t]
%	\vspace{-2mm}
	\begin{minipage}[b]{\columnwidth}
\begin{lstlisting}[escapechar=^,label=lst:old_homogeneous_indexing,basicstyle=\ttfamily\small, caption=Processing of multiple bitmaps with 1-bit entries.]
static final int mask(int hash, int shift) {^\label{func:mask}^
  return (hash >>> shift) & 0b11111; 
}^\label{func:mask-end}^
static final int index(int bitmap, int bitpos) {^\label{func:index}^
  return Integer.bitCount(bitmap & (bitpos - 1));
}^\label{func:index-end}^

// processing in (Heterogeneous) CHAMP
void processAtNode(int keyHash, int shift) {^\label{func:processAtNode}^
  int mask = mask(keyHash, shift);
  int bitpos = bitpos(mask);
  
  int nodeMap = nodeMap();
  if ((nodeMap & bitpos) != 0) {
    // process node
    int index = index(nodeMap, bitpos);
    ^\etcetera^
  } else {
    int dataMap = dataMap();
    if ((dataMap & bitpos) != 0) {
      // process payload category 1
      int index = index(dataMap, bitpos);
      ^\etcetera^
    } else {
      int xxxxMap = xxxxMap();
      if ((xxxxMap & bitpos) != 0) {
        // process payload category X
        int index = index(xxxxMap, bitpos);
        ^\etcetera^
      } else {
        // process empty slot
        ^\etcetera^
    }
  }
}
\end{lstlisting}
	\end{minipage}
	\hfill
	\begin{minipage}[b]{\columnwidth}
\begin{lstlisting}[escapechar=^,label=lst:new_heterogeneous_indexing, basicstyle=\ttfamily\small, caption=Processing of one bitmaps with 2-bit entries.]
static final int index(long bitmap, long bitpos) {
  return Long.bitCount(bitmap & (bitpos - 1));
}

// processing in a Heterogeneous HAMT
void processAtNode(int keyHash, int shift) {
  long bitmap = bitmap();
  
  int mask = mask(keyHash, shift) << 1;
  long bitpos = 1L << mask;^\label{snippet:extracting-type}^
  
  int pattern = (int) ((bitmap >>> mask) & 0b11);^\label{var:pattern}^
  Type type = toEnum(pattern);^\label{snippet:extracting-type-end}^
  
  switch (type) {
  case EMPTY:
    ^\etcetera^    
    break;
  case NODE:
    int index = index(filter(bitmap, type), bitpos);
    ^\etcetera^
    break;
  case PAYLOAD_CATEGORY_1:
    int index = index(filter(bitmap, type), bitpos);
    ^\etcetera^
    break;
  case PAYLOAD_CATEGORY_2:
    int index = index(filter(bitmap, type), bitpos);  
    ^\etcetera^
    break; 
  }
}
\end{lstlisting}	
	\end{minipage}
	\vspace{-1mm}
\end{figure*}
The heterogeneous skeleton in Listing~\ref{lst:skeleton}~(lines \ref{HHAMTstart}--\ref{HHAMTend}) does not exhibit an optimal encoding. We specialize the \lstinline{BitVector} code for obtaining better memory performance. Assuming $k=2$, we use a single \lstinline{long} field as bitmap, for a larger $k$ we would use several consecutive \lstinline{int} or \lstinline{long} fields. 

The way we index into the trie node array (for lookup, insertion or deletion) is a key design element. This indexing is different between the original \CHAMP encoding and the new \HHAMT encoding because there are $k$-cases to distinguish.

Listing~\ref{lst:old_homogeneous_indexing} shows how \CHAMP's original per-node-bitmap indexing would work if generalized to multiple entry types. By default \CHAMP already distinguishes between payload data and nested nodes with separate bitmaps. This baseline (naive) design for heterogeneous hash tries carries on similarly to distinguish more types of references. The masking function (lines~\ref{func:mask}--\ref{func:mask-end}) selects the prefix bits based on the node level in the tree (\lstinline{shift = 5 * level}). The index function (line~\ref{func:index}--\ref{func:index-end}) requires a \lstinline{bitpos} variable with a single non-zero bit, designating one of the \num{32} possible branches. It then maps from the \lstinline{bitmap}/\lstinline{bitpos} tuple to a sparse-array index by counting the non-zero bits in \lstinline{bitmap} on the right of \lstinline{bitpos}. On line \ref{func:processAtNode} a method template for lookup, insertion, and deletion is shown. Because for each of the three data categories a separate bitmap is used the processing happens in a linear-scanning manner until the right category for a hash-prefix is matched, or the default case applies (line~31). 

Although lines~12, 18, and 24 suggest the use of separate bitmaps for each distinct type, two bitmaps are sufficient to distinguish between three cases:
\begin{lstlisting}
int xxxxMap = rawMap1 & rawMap2;
int dataMap = rawMap2 ^ xxxxMap;
int nodeMap = rawMap1 ^ xxxxMap;
\end{lstlisting}
The above listing depicts how to retrofit three logical bitmaps onto two physical bitmaps. The fields for \texttt{datamap} and \texttt{nodemap} are renamed to \texttt{rawMap1} and \texttt{rawMap2}. Subsequently, the data structure infers three logical views from the two raw bitmaps. We further will refer to this retrofitted heterogeneous variant as \HCHAMP. 

\begin{figure}[t]
	\vspace{-2mm}
	\begin{lstlisting}[label=lst:new_heterogeneous_filter, basicstyle=\ttfamily\small, caption={Filtering of multi-bit patterns (for $k = 2$).}]
static final long filter(long bitmap, Type type) {
  long mask = 0x5555555555555555L;

  long masked0 = mask & bitmap;
  long masked1 = mask & (bitmap >> 1);

  switch (type) {
  case EMPTY:
    return (masked0 ^ mask) & (masked1 ^ mask);
  case NODE:
    return masked0 & (masked1 ^ mask);
  case PAYLOAD_CATEGORY_1:
    return masked1 & (masked0 ^ mask);
  case PAYLOAD_CATEGORY_2:
    return masked0 & masked1;
  }
}
	\end{lstlisting}
\end{figure}

Listing~\ref{lst:new_heterogeneous_indexing} illustrates operations on the bitmap in the generalized data structure that is specialized  to $k=2$. 
The mask function can be reused, and the \lstinline{index} function is scaled to using a \lstinline{long}. 
The new template method retrieves the 2-bit wide \lstinline{pattern} (line~\ref{var:pattern}) and translates it to an enum value to switch on. Instead of having to search linearly, as in Listing~\ref{lst:old_homogeneous_indexing}, we now jump directly to the relevant case handler. Using a fast \lstinline{switch} is even more beneficial with an increasing number of heterogeneous types ($k > 2$), and while iterating which is when type dispatch will be hot. 

\subsection{Optimizing Bit-Counting} 

Extra bitwise operations are in the overhead of \HHAMT which we need to mitigate. We explain three techniques to do so.

\paragraph{Relative Indexing into a Single Data Category.} The purpose of the \lstinline{index} function in Listing~\ref{lst:new_heterogeneous_indexing} is to calculate the relative index of a data element within its data category. Given a \lstinline{type} enum and a trie-branch descriptor (\lstinline{bitpos}), the \lstinline{index} function calculates how often the given \lstinline{type} pattern occurs in the bitmap before the \lstinline{bitpos} position. 

The Java standard library contains bit count operations for the types \lstinline{int} and \lstinline{long} that count the number of bits set to 1. These functions do not support $n$-bit patterns with $n > 1$. However, we want to reuse the aforementioned functions, because on the widespread \emph{X86/X86\_64} architectures they map directly to hardware instructions. We introduce some bitmap pre-processing with filters to get to that point where we can use the native bit counters. Listing~\ref{lst:new_heterogeneous_filter} illustrates how such a filter reduces a matching 2-bit wide pattern to a single bit set to 1, while resetting all other bits to 0. 

\paragraph{Distribution of Heterogeneous Elements.} While lookup, insertion, and deletion only require indexing into a single data category, on the other hand iteration and streaming require information about the types of all elements in a trie node: their frequency per node. Studies on homogeneous data structures~\cite{Bolz:2013ku} have shown avoiding checks on a per elements basis is indeed relevant for performance. 

To also avoid such checks in \HHAMT we introduce the use of histograms, on a per node basis, that are calculated in constant time (for a given branch factor). The computation is independent of the number of heterogenous types: 
\begin{lstlisting}[escapechar=!]
int[] histogram = new int[!$2^n$!];

for (int branch = 0; branch < 32; branch++) {
  histogram[(int) bitmap & mask]++;
  bitmap = bitmap >>> n;
}
\end{lstlisting}
The former listing abstracts over the number of heterogeneous elements and executes in 32 iterations. \lstinline{n} and \lstinline{mask} are constants, where \lstinline{mask} has the lowest $n$ bits set to 1. In its generic form, the code profits from default compiler-level optimizations ---such as scalar replacement~\cite{partical-escape-analysis-stadler-thesis} to avoid allocating the array on the heap, and loop unrolling. 

We assigned the bit-pattern \lstinline{EMPTY = 0} and \lstinline[mathescape=true]{NODE = $2^{n-1}$},\break the various remaining heterogenous types are assigned consecutively in the middle. For iteration, streaming, or batch-processing data, histograms avoid expensive repetition of indexing individual categories: $k$ bit-count operations, where each one requires applying a filter to the bitmap.
For example, the total number of elements, regardless of their types, can be calculated with \lstinline{32 - histogram[EMPTY] - histogram[NODE]}. The otherwise complex code for trie-node iteration reduces to looping through the two-dimensional histogram using two integer indices. The added benefit is that inlined values although stored out of order, will be iterated over in concert, avoiding spurious recursive traversal and its associated cache misses~\cite{Steindorfer:2015iz}. Finally, iteration can exit early when the running counter reaches the maximum branching factor of 32 to avoid iterating over empty positions in the tail.
Note that for fixed $k$ the code can be partially evaluated (i.e., by a code generator) to avoid the intermediate histogram completely.

\paragraph{Reversing the Bitmap Encoding: Extracting Index and Type.} For enabling fast structural equality comparisons~\cite{Steindorfer:2015iz} maintaining a canonical form of the hash-trie is essential, also after the delete operation. For \HHAMT and especially for \HHAMT multi-maps this takes an extra effort: the deletion operation does know the index and type of the removed element, however it does not know the index and type of the remaining elements. Upon deletion, canonicalization triggers inlining of sub-tree structures with only a single remaining payload tuple. Efficiently recovering the index an type of the only remaining tuple is important for the overall efficiency of the deletion operation. We devised a recovery function for bitmaps with $n$-bit tuples, based on Java's standard library functions: \lstinline{Long.numberOfTrailingZeros(bitmap)/n*n}. By first counting the number of trailing zeros, we approximate the region within the bitmap that contains bit-pattern information. We subsequently adjust the non-zero count to our $n$-bit pattern alignment with an integer division followed by an multiplication. As a result, we recovered the \lstinline{mask} that allows retrieving the type of the remaining element (cf. Listing~\ref{lst:new_heterogeneous_indexing}, lines \ref{snippet:extracting-type}--\ref{snippet:extracting-type-end}).

\paragraph{Outlook.} We have now discussed all techniques to mitigate \CPU overhead caused by a more complex indexing. The remaining challenge is saving memory, which is discussed next.
% !TEX encoding = UTF-8 Unicode
% !TEX root = ../paper.tex

\section{Lean Specializations\label{sec:conribution_scalable_specialization_encoding}}

Specialization for a fixed number of heterogeneous types will prove essential for both memory efficiency and \CPU performance. In this section we take the perspective of the general $k$-heterogeneous \HHAMT. The effect of these techniques will be evaluated in Sections~\ref{sec:evaluation-internal}, \ref{sec:evaluation-external}, \ref{sec:evaluation-primitive-collections} and \ref{sec:evaluation-real-world} in different contexts.

For a \HHAMT with $k$ different types, there exist $arity_{nodes} \times \prod_{i = 1}^{k} arity_i$ possible strongly-typed variants in theory, with the constraint that $arity_{nodes} + \sum_{i = 1}^{k} arity_i <= 32$. We can reduce this complexity by grouping different heterogeneous types together into a section that is represented by their least upper bound type. Ultimately, we can group together all reference types and sub-nodes into one section, and all primitive types into another section~\cite{Ureche:2013gj}, to achieve a quadratic upper bound that overcomes the dichotomy of reference and primitive types. Therefore, in the remainder of this section we will focus on the most common case of $k=2$ that also satisfies our use case of multi-maps. Note that due to the bitmap encoding we always know the precise type of an object, using more general types for internal storage is solely used to reduce the total number of specializations. 

There exist empirical evidence~\cite{Steindorfer:2014dk} for $k=1$ that specializing up to arities of $8$ or $12$ balances impact on memory performance best with the necessary amount of generated code. However with heterogeneity $k > 1$ this may not hold, and to exploit inlining primitive types for saving more memory we should support specializing the full bandwidth up to 32. 

We now present an improved approach for code generation that allows fully specialized collections (i.e., "array-less" data structures) with very low memory footprints. It aims to overcome the following issues that typically compromise performance of specialized code:
\begin{description}
\item[Additional Polymorphism:] Turning a generic data type into a set of distinct specializations compromises trace-based inlining strategies of a \JIT compiler. By introducing specializations, previous monomorphic call-sites are turned into polymorphic call-sites. Thus a \JIT compiler has to fallback to dynamic dispatch for method method calls that were previously resolved to direct calls.
\item[Code Bloat:] Substituting a dynamic structure with specialiations often demands the specialization of operations as well. In the case of hash-tries, we specialize for constant array sizes~\cite{Steindorfer:2014dk}: instead of referencing a heap structure, we inline the array-fields into a trie-node. Unfortunately the resulting memory saving come at a price: suddenly array operations (i.e., allocation, copy, get, set, length) must be specialized as well. 
\item[Interdependency of Specializations:] In general, each specialized data type contains static references to other specializations that represent possible next states. Listing~\ref{lst:specialization_interlinking} exemplary lists the \lstinline{add} method of set data structure specialized for one element that might return a set specialized for two elements. The switching between specialized representations, puts strain on the \JIT compiler at run-time due to incremental class loading and the constant need to compile methods of specializations during a data structure builds up, further, it is one source of code bloat.
\end{description}
\begin{figure}[t]
	\vspace{-2mm}
	\begin{lstlisting}[escapechar=!, label=lst:specialization_interlinking, caption=Interlinking of specializations prohibits generic methods: \lstinline{Set1} contains a static reference to \lstinline{Set2}.]
abstract class Set1 implements Set {
	final Object slot0;

	Set add(Object value) {
		if (slot0.equals(value)) {
			return this;
		} else {
			return new Set2(slot0, value);
		}
	}
}
	\end{lstlisting}
\end{figure}

\noindent In the remainder of this section we detail our approach of specialization that remedies the aforementioned overheads. In our design, a specialization represents purely a heterogeneous trie-node, specialized for a certain content size. It contains pre-evaluated content stored in static fields and instance fields (storing the bitmap, and the inlined array content), however does not override methods. 

\subsection{Indexing and Selecting Specializations}

We replace the use of arrays, which can be allocated using an arbitrary length parameter, with fields inlined in specialized classes. Commonly, for each specilization a unique constructor must be called (cf. Listing~\ref{lst:specialization_interlinking}, specialization interlinking). Which constructor must be called depends on the current state of a trie node and the operation applied to the data structure. % (e.g, insertion or deletion). 

To enable class selection at run-time, we introduce a global and static two-dimensional \lstinline{Class[][] specializations} array, indexed by the number of primitive data fields ($t$) and the number of reference fields ($u$). This lookup table solves the interdependency problem of specialization: when adding a key-value tuple of reference type the next specialization can be be determined with \lstinline{specializations[t][u + 2]}, or respectively with \lstinline{specializations[t - 2][u]} when a tuple of primitive type is deleted. Once a specialization is selected, it can be initialized by invoking its default constructor: \lstinline{Object instance = specialization[t][u].newInstance()}. 

Since the array is often used and relatively small, we found it runs faster than distributing code over the specialized classes. This also allows for more generic code in base classes which is therefore used more often and more likely to be optimized by the \JIT compiler.

\subsection{Initializing Instances of Specialized Classes} 

For the generic representation that operates on arrays, we would use \lstinline{System.arraycopy} initializing a new trie node, which is really fast. Now we want to try and approach similar efficiency for initializing the fields of our specialized classes. 

Our solution is to introduce a \lstinline{arraycopy}-like operation that is capable of copying consecutive fields between object instances: an \lstinline{ArrayView} on an object layout is an abstraction which logically maps an arbitrary region within objects to an array. To ensure safety we check whether the \JVM indeed maps the fields in a consecutive region at class loading time. Using a primitive \lstinline{ArrayView.copy} we achieve similar performance to \lstinline{System.arraycopy}. We measured the effect using a micro-experiment: the new primitive is about \SIrange{20}{30}{\percent} faster than field-by-field copying. Since eventually copying trie nodes is the primary bottleneck we may expect around similar speedups for insertion- and deletion-intensive code of \HHAMT and less for lookup intensive code.

Listing~\ref{lst:heap_region_array_view} shows how we can model an array view of a range of fields within a heap object. Once we have obtained a reference to an \lstinline{ArrayView}, we can invoke corresponding \lstinline{(getFrom|setIn)HeapRegionArrayView} methods that either retrieve or a set a value of a \lstinline{ArrayView}. To mimic \lstinline{System.arraycopy} on an \lstinline{ArrayView}, we use \lstinline{sun.misc.Unsafe.} \lstinline{copyMemory}. For our experiments, we extended the \lstinline{copyMemory} function to support copying from/to objects while catering for card marking, i.e., signaling the \GC that references changed.

\begin{figure}[t]
	\vspace{-2mm}
	\centering
	\begin{lstlisting}[escapechar=!, label=lst:heap_region_array_view, caption=\lstinline{ArrayView} on regions of specialized trie node.]
class TrieNode_T4_U3 implements Node {
  long bitmap;

  int key0; int val0;
  int key1; int val1;
  
  Object slot0;
  Object slot1;
  Object slot2;
  
  static ArrayView getArrayView_T4() {
    return createHeapRegionArrayView(
          TrieNode_T4_U3.class, "key0", "val1");
  }

  static ArrayView getArrayView_U3() {
    return createHeapRegionArrayView(
          TrieNode_T4_U3.class, "slot0", "slot2");
  }    
}
	\end{lstlisting}
\end{figure}

\paragraph{Relationship to \lstinline{VarHandle} \API of the Upcoming \acs{JDK} 9.} The \JDK 9 will introduce an \API for uniformly referencing, accessing and modifying fields. Thus, independently of the realization of a variable ---static field, instance field, or array--- a handle to the field reference can be obtained. In earlier versions of Java, the granularity of references was restricted to objects; a \lstinline{VarHandle} in contrast enables addressing fields or arrays (at a finer granularity) inside an object. The \lstinline{VarHandle} \API furthermore contains abstractions to view and process off-heap memory regions as arrays. However, it does not provide likewise abstractions for obtaining array views on on-heap regions. 

The aforementioned \lstinline{ArrayView} implementation we used provides a proof-of-concept implementation on how to extend the \lstinline{VarHandle} \API to support array views for on-heap regions.

\subsection{Summary} 

In the context of collections, we eliminated issues that typically compromise the performance of specialized code. We will evaluate the effects of these techniques in Section~\ref{sec:evaluation-internal}.
%\input{chapters/efficient_data_access.tex}
% !TEX encoding = UTF-8 Unicode
% !TEX root = ../paper.tex

\section{Assessing the Cost of Multi-Maps\label{sec:evaluation-internal}}

\newcommand{\PROBESIZE}{\num{8}\xspace}

\newcommand{\bmLookup}{Lookup\xspace}
\newcommand{\bmInsert}{Insert\xspace}
\newcommand{\bmDelete}{Delete\xspace}
\newcommand{\bmLookupFail}{Lookup~(Fail)\xspace}
\newcommand{\bmInsertFail}{Insert~(Fail)\xspace}
\newcommand{\bmDeleteFail}{Delete~(Fail)\xspace}
\newcommand{\bmIterationKey}{Iteration~(Key)\xspace}
\newcommand{\bmIterationEntry}{Iteration~(Entry)\xspace}
\newcommand{\bmEqualityDistinct}{Equality~(Distinct)\xspace}
\newcommand{\bmEqualityDerived}{Equality~(Derived)\xspace}
\newcommand{\bmEqualityDifferent}{Equality~(Different)\xspace}
\newcommand{\bmFootprintThreeTwo}{Footprint~(\num{32}-bit)\xspace}
\newcommand{\bmFootprintSixFour}{Footprint~(\num{64}-bit)\xspace}

In this section we evaluate the performance characteristics of the various implementations of multi-maps on top of the \HCHAMP, \HHAMT, and specialized \HHAMT encodings, comparing them against the basic homogeneous \CHAMP map data structure as state-of-the-art baseline~\cite{Steindorfer:2015iz}. We are interested in isolating the effects that are incurred by adding the heterogeneity feature: 
\begin{itemize}
\item \HCHAMP is close to \CHAMP (same logic, but derives three bitmap views from two physical bitmaps); 
\item \HHAMT generalizes the heterogeneous bitmap encoding; 
\item and specialized \HHAMT improves memory footprints by dynamically selecting statically known specializations. 
\end{itemize}

In the case of multi-maps, heterogeneity lies in the internal distinction between $1 : 1$ and $1 : n$ mappings.

\paragraph{Assumptions.} We evaluate the pure overhead of operations on the data structures, without considering cost functions for \texttt{hashCode} and \texttt{equals} methods. This performance assessment should reveal the overhead of adding heterogeneity to \CHAMP, the effect of the specialization approach and the effect of accessing the heterogeneous data elements.

\paragraph{Hypotheses.} We expect \HHAMT's runtime performance of lookup, deletion, and insertion to be similar comparable to \CHAMP's runtime performance, but never better. Running times should not degrade below a certain threshold ---we feel that \SI{25}{\percent} for median values and \SI{50}{\percent} for maximum values would about be acceptable as a trade-off--- (Hypothesis 1).

Iteration over a multi-map is more complex than iterating over a map. \bmIterationKey has to distinguish between heterogeneous categories, whereas \bmIterationKey has to distinguish heterogeneous categories, \bmIterationEntry additionally has to flatten nested sets to obtain a tuple view on multi-maps. Consequently, we assume costs of about \SI{25}{\percent} for median values and \SI{50}{\percent} for maximum values as well (Hypothesis 2). 

Based on related work in the domain of specializing \acp{HAMT}~\cite{Steindorfer:2014dk}, we expect that specializing may introduce run-time overhead. However, we expect lower overhead (than the reported \SIrange{20}{40}{\percent} degradations for lookup) due to our mitigation strategies outlined in Section~\ref{sec:conribution_scalable_specialization_encoding} (Hypothesis 3). 

Furthermore, memory footprints of \HCHAMP and \HHAMT should in practice match \CHAMP's footprints, because all variants use in total 64-bits for bitmaps (Hypothesis 4).

\subsection{Experiment Setup\label{sec:experiment-setup-internal}}

We use a machine with Apple OS X (10.11.3) and \SI{16}{\giga\byte} RAM. It has an Intel Core i7-3720QM CPU, with \SI{2.60}{\giga\hertz}, and an \SI{6}{\mega\byte} \LLC. Frequency scaling was disabled. 
For testing, we used an OpenJDK (\acs{JDK} 8u65) \JVM configured with a fixed heap size of \SI{8}{\giga\byte}.
We measure the exact memory footprints of data structures with Google's memory-measurer library.\footnote{\url{https://github.com/DimitrisAndreou/memory-measurer}} 
Running times of operations are measured with the \JMH, a framework to overcome the pitfalls of microbenchmarking.\footnote{\url{http://openjdk.java.net/projects/code-tools/jmh/}} 
For all experiments we configured \JMH to perform 20 measurement iterations of one second each, after a warmup period of 10 equally long iterations. For each iteration we report the median runtime, and measurement error as \MAD{}, a robust statistical measure of variability that is resilient to small numbers of outliers. Furthermore, we configured \JMH to run the \GC between measurement iterations to reduce a possible confounding effect of the \GC on time measurements.

In our evaluation we use collections of sizes $2^x$ for $x \in [1,23]$. Our selected size range was previously used to measure the performance of \acp{HAMT}~\cite{Bagwell:2001tw,Steindorfer:2015iz}.
For every size, we fill the collections with numbers from a random number generator and measure the resulting memory footprints. Subsequently we perform the following operations and measure their running times:
\begin{description}
\item[\bmLookup, \bmInsert and \bmDelete:] Each operation is measured with a sequence of \PROBESIZE random parameters to exercise different trie paths. For \bmLookup and \bmDelete we randomly selected from the elements that were present in the data structures.\footnote{For $<$ \PROBESIZE elements, we duplicated the elements until we reached \PROBESIZE samples.} For \bmInsert we ensured that the random sequence of values was not yet present.
\item[\bmLookupFail, \bmInsertFail and \bmDeleteFail:] Measuring unsuccessful operations. The setup equals the aforementioned setting, however with the difference that we swap the sequences of present/not present parameters.
\item[\bmIterationKey:] Iterating over the elements of a set or the keys of a map respectively.
\item[\bmIterationEntry:] Iterating over a multi-map, flattening and yielding tuples of type \texttt{Map.Entry}.
\end{description}

We repeat the list of operations for each size with five different trees, starting from different seeds. This counters possible biases introduced by the accidental shape of the tries, and accidental bad locations in main memory. Evaluating \HAMT data structures containing simply random integers accurately simulates any application for which the elements have good uniformly distributed hash codes. A worse-than-uniform distribution would ---regardless of the \HAMT-like implementation--- overall reduce the memory overhead per element and increase the cost of updates (both due to clustering of elements). We consider a uniform distribution the most representative choice for our comparison.

\subsection{Experiment Results\label{sec:results-runtime}}

\begin{figure}[t]%
	\centering%
	\subfloat[\HCHAMP multi-map versus \CHAMP map (baseline).]{%
		\includegraphics[width=\columnwidth]{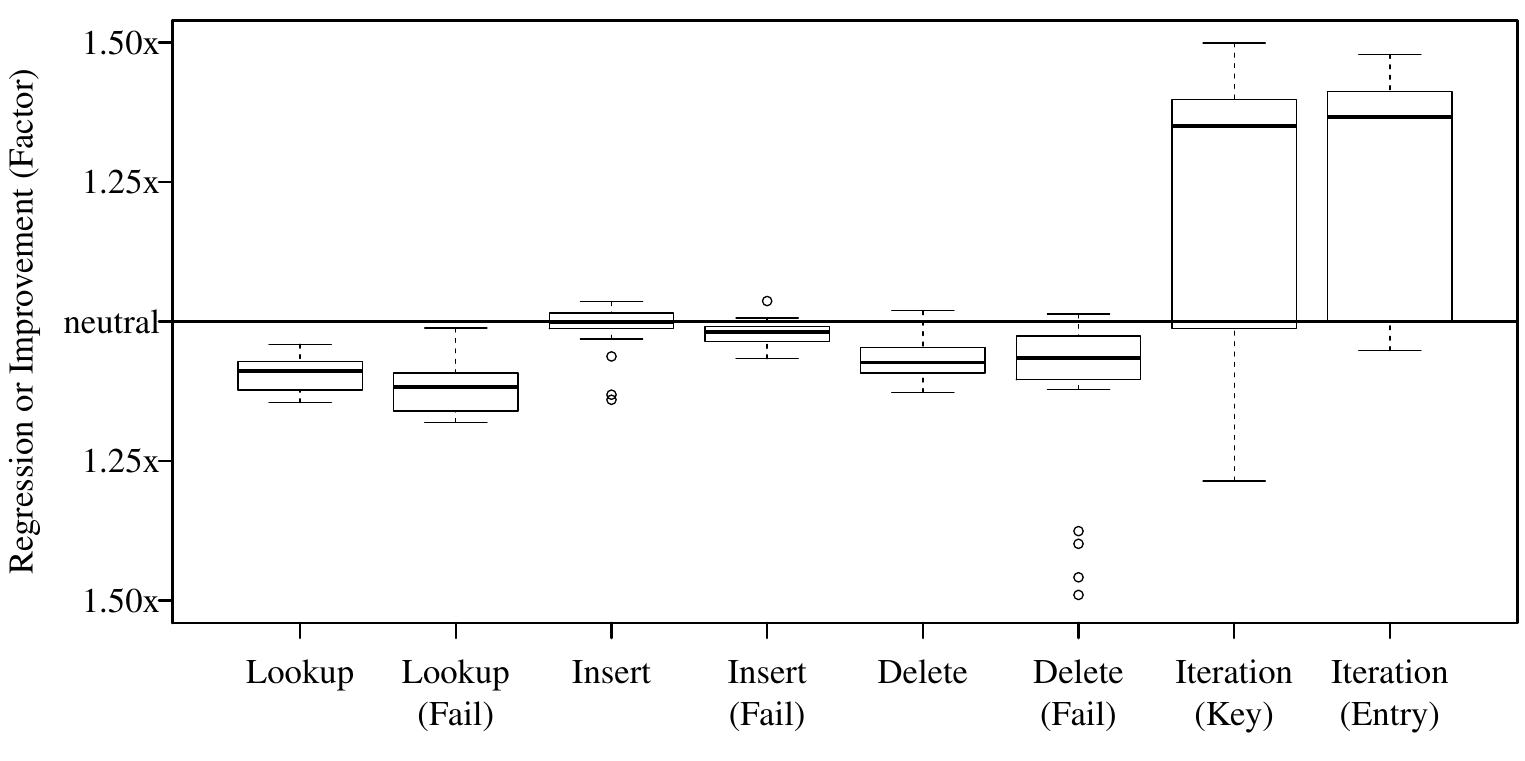}\label{fig:benchmarks-champ-map-vs-hchamp-multimap-boxplot}%
	}%
	
	\subfloat[\HHAMT multi-map versus \CHAMP map (baseline).]{%
		\includegraphics[width=\columnwidth]{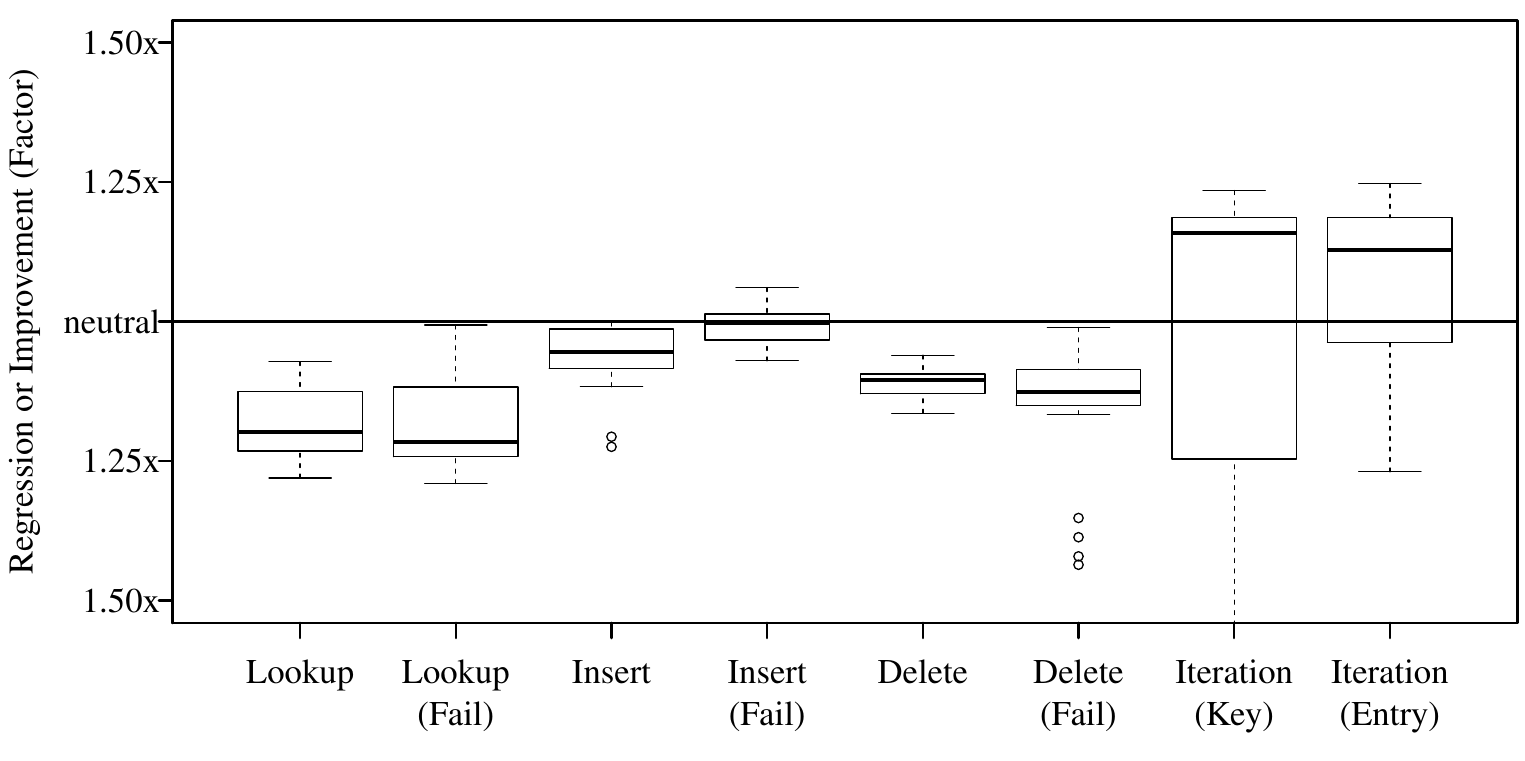}\label{fig:benchmarks-champ-map-vs-hhamt-multimap-boxplot}%
	}%
	\vspace{3mm}%
	\caption{Visualizing the overhead of various multi-map implementations over a \CHAMP map implementation.\label{fig:benchmarks-champ-map-vs-others}}	
\end{figure}

We first report the precision of the individual data points. For \SI{99}{\percent} of the data points, the relative measurement error amounts to less than \SI{1}{\percent} of the microbenchmark runtimes, with an overall range of \SIrange{0}{4.8}{\percent} and a median error of \SI{0}{\percent}.

We summarize the data points of the runs with the five different trees with their medians. Then Figure~\ref{fig:benchmarks-champ-map-vs-hchamp-multimap-boxplot}, and \ref{fig:benchmarks-champ-map-vs-hhamt-multimap-boxplot} report for each benchmark the ranges of runtime improvements or degradations. For brevity, the effects on memory footprints and of specialization are not contained in the boxplots, but are discussed in text. Each boxplot visualizes the measurements for the whole range of input $\operatorname{size}$ parameters. For improvements we report speedup factors above the neutral line ($\operatorname{measurement}_{\operatorname{CHAMP}}/\operatorname{measurement}_{\operatorname{HHAMT-Variant}}$), and degradations as slowdown factors below the neutral line , i.e., the inverse of the speedup equation.
From this data we learn the following:
\begin{description}
\item [Confirmation of Hypothesis 1:] The cost of converting a map to a multi-map stayed within the specified bounds for both \HCHAMP and \HHAMT. 

For \HCHAMP, \bmLookup, \bmInsert and \bmDelete added a median slowdown of \SI{9}{\percent}, \SI{0}{\percent}, and \SI{7}{\percent} respectively, and \bmLookupFail, \bmInsertFail and \bmDeleteFail added \SI{12}{\percent}, \SI{2}{\percent} and \SI{7}{\percent} respectively. With exception to single outliers produced \bmDeleteFail, the maximum slowdown are lower than \SI{18}{\percent} at most.

For the generalized \HHAMT, the costs for multi-maps over maps are higher. \bmLookup, \bmInsert and \bmDelete added a median slowdown of \SI{20}{\percent}, \SI{5}{\percent}, and \SI{10}{\percent} respectively, and \bmLookupFail, \bmInsertFail and \bmDeleteFail added \SI{22}{\percent}, \SI{0}{\percent} and \SI{13}{\percent} respectively. With exception to single outliers produced \bmDeleteFail, the maximum slowdown are lower than \SI{29}{\percent} at most.

\item [(Partial) Confirmation of Hypothesis 2:] Compared to our baseline, and counter to our intuition, \HCHAMP improved \bmIterationKey by a median \SI{35}{\percent} and \bmIterationEntry by \SI{37}{\percent}. The more general \HHAMT improved \bmIterationKey by a median \SI{16}{\percent} and \bmIterationEntry by \SI{13}{\percent}. However, according to Figure~\ref{fig:benchmarks-champ-map-vs-others}, the value spread appears large and the maximum bounds are violated for \bmIterationKey.

\item [(Partial) Confirmation of Hypothesis 3:] On average, we observed an overhead of \SI{3}{\percent} for \bmLookup and \SI{6}{\percent} for \bmLookupFail when comparing a specialized \HHAMT against its regular \HHAMT counterpart. These numbers confirm our intuition and are lower then the \SIrange{20}{40}{\percent} overhead reported by Steindorfer and Vinju~\cite{Steindorfer:2014dk}. The median costs for \bmInsert (\SI{24}{\percent}) and \bmDelete (\SI{31}{\percent}) however match their results. Concerning memory consumption, specializations improved memory consumption by at least \SI{38}{\percent} for data structures with 32 or more entries. 

\item [Confirmation of Hypothesis 4:] \sloppypar Memory footprints of \HCHAMP and \HHAMT (omitted in Figure~\ref{fig:benchmarks-champ-map-vs-others}) match exactly the footprint of \CHAMP, when using multi-maps as maps. 
\end{description}

\paragraph{Discussion.}
A more detailed investigation revealed that for \bmIterationKey measurements at sizes $2^1$ and $2^5$ showed significant deviation from the remaining measurements. These two measurements were not statistically identified as outliers due to the small sample size of 23 (sizes $2^x$ for $x \in [1,23]$). When removing these two measurements, the upper bound of slowdowns is \SI{6}{\percent} for \HHAMT and \SI{36}{\percent} for \HCHAMP.

While not impacting lookup performance, specializing trades the runtime performance of insertion and deletion for gaining savings of approximately \SI{1.4}{\factor}.\footnote{Note that only the outer multi-map structure was specialized and not the nested sets. A further specialization of the nested sets would yield even more substantial memory savings.} Because only operations that allocate new tree nodes are affected, we attribute slowdowns to the lookup table we introduced (adding two memory indirection). Nevertheless, specializing is of key importance when optimizations for primitive data types; we evaluate that effect separately in Section~\ref{sec:evaluation-primitive-collections}.

\paragraph{Summary.}
Despite its more complex and heterogeneous encoding, \acp{HHAMT} achieves excellent runtimes across all tested operations. Converting a map into a multi-map with the means of a heterogeneous encoding had usually less costs associated than we expected beforehand. Our specialization approach could successfully mitigate overhead for lookups while reducing memory footprints. However, using a lookup table for our specializations still impacts insertion and deletion, when compared to regular array allocations that do not require a lookup table.

\section{Comparing Immutable Multi-Maps\label{sec:evaluation-external}}

\begin{figure}[t]%
	\centering%
	\subfloat[Specialized \HHAMT multi-map versus Clojure's multi-map (baseline).]{%
		\includegraphics[width=\columnwidth]{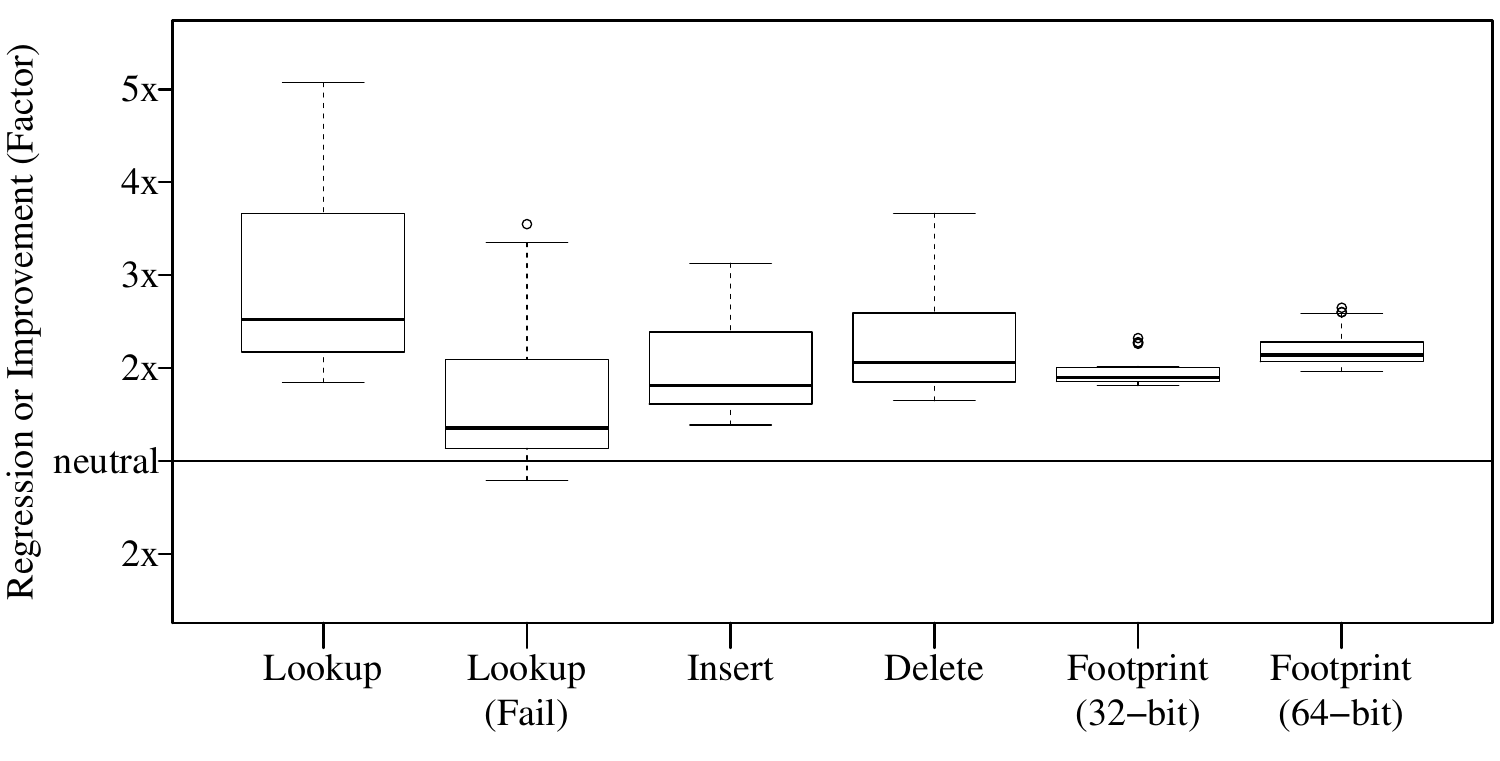}\label{fig:all-benchmarks-clojure-multimap-boxplot}%
	}%
	\qquad%
	\subfloat[Specialized \HHAMT multi-map versus Scala's multi-map (baseline).]{%
		\includegraphics[width=\columnwidth]{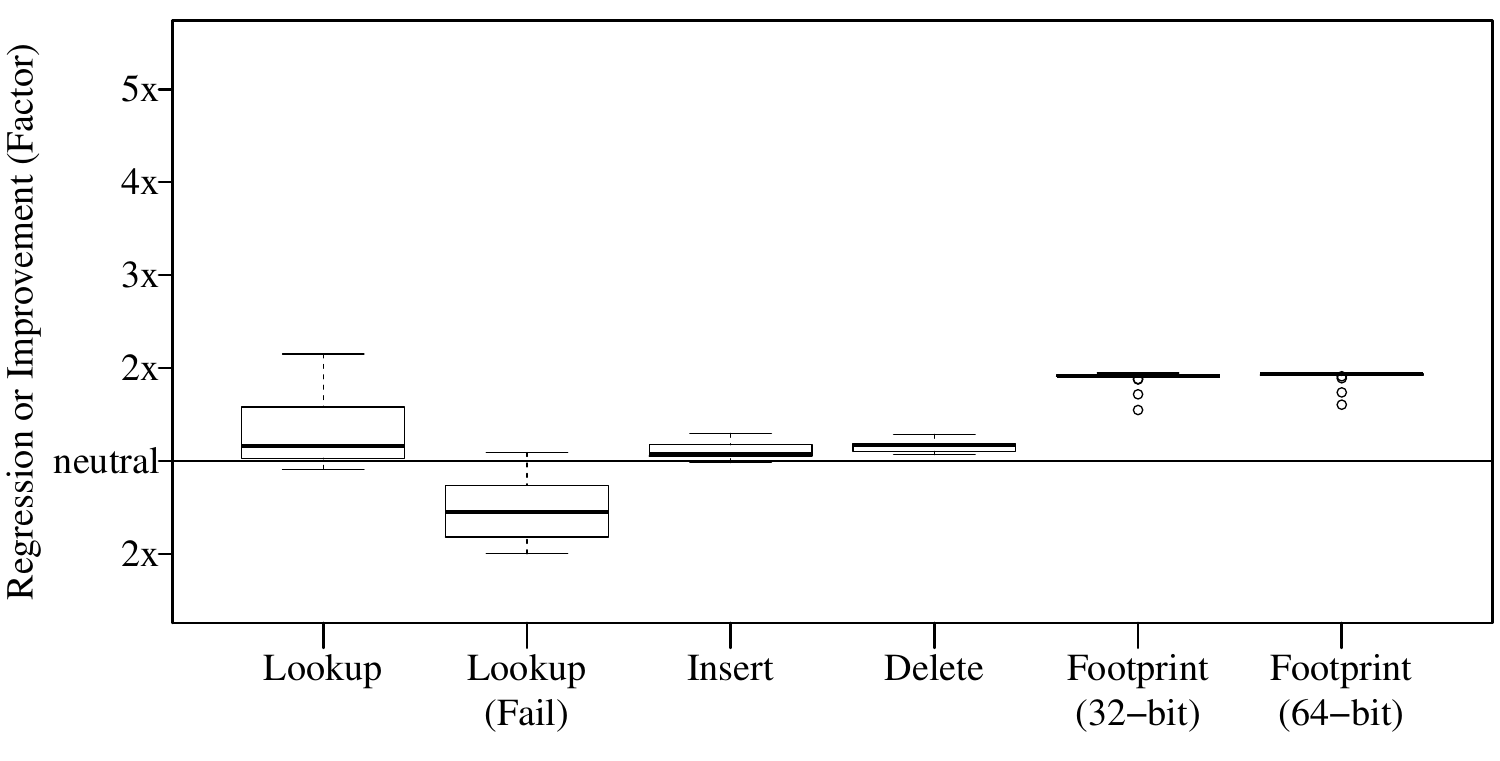}\label{fig:all-benchmarks-scala-multimap-boxplot}%
	}%	
	\vspace{3mm}%
	\caption{Performance comparison of a specialized \HHAMT multi-map against implementations in Clojure and Scala.}	
\end{figure}

We further evaluate the performance characteristics of our specialized \HHAMT multi-map against implementations from Clojure and Scala. Both languages do not provide native immutable multi-maps in their standard libraries, however suggest idiomatic solutions to transform maps with nested sets into multi-maps. 

VanderHart~\cite[p.~100--103]{vanderhart2014clojure} proposes a solution for Clojure based on ``protocols''. Values are stored untyped as either a singleton, or a nested set. Consequently, the protocol extension handles the possible case distinctions ---not found, singleton, or nested set--- for lookup, insertion, and deletion. 

Scala programmers would idiomatically use a trait for hoisting a regular map to a multi-map. However, the Scala standard library only contains a trait for mutable maps; we therefore ported the standard library program logic of the trait to the immutable case, nesting typed sets into maps. 

\paragraph{Hypotheses.} We expect specialized \HHAMT's runtime performance of lookup, deletion, and insertion to equal the competitors performance, because we tried hard to mitigate the incurred overhead, and the idiomatic solutions require some overhead as well. Runtimes should not degrade below a certain threshold ---say \SI{10}{\percent} for median values and \SI{20}{\percent} for maximum values would just be acceptable--- (Hypothesis 5).
However, for negative lookups we expected that specialized \HHAMT performs worse than Scala (Hypothesis 6). This hypothesis is based on related work~\cite{Steindorfer:2015iz} that explains the inherent differences between \CHAMP and Scala when it comes to memoizing hash codes. Our hypothesis expects memory improvements by at least \SI{50}{\percent} on average due to omitting nested collections for singletons (Hypothesis 7). 

\subsection{Experiment Setup}

Data generation is derived from the experimental setup outlined in Section~\ref{sec:experiment-setup-internal}. We keep the number of unique keys equal ---$2^x$ for $x \in [1,23]$--- but instead of using distinct data in each tuple, we now use \SI{50}{\percent} of $1 : 1$ mappings, and \SI{50}{\percent} of $1 : 2$ mappings. 
Fixing the maximal size of right-hand side of the mapping to $2$ may seem artificial, but it allows us to precisely observe  the singleton case, the case for introducing the wrapper and the overhead per additionally stored element. The effect of larger value sets on memory usage and time can be inferred from that without the need for additional experiments.

\begin{description}
\item[\bmInsert:] We call insertion in three bursts, each time with \PROBESIZE random parameters to exercise different trie paths. Firstly we provoke full matches (key and value present), secondly partial matches (only key present), and thirdly no matches (neither key nor value present). Next to insertion of a new key, this mixed workload also triggers promotions from singletons to full collections.
\item[\bmDelete:] We call deletion in two bursts, each time with \PROBESIZE random parameters. Provoking again, full matches and partial matches. Next to deletion of a key, this mixed workload also triggers demotions from full collections to singletons, and canonicalization where applicable.
\item[\bmLookup:] Similar to \bmDelete we call lookup in two bursts to exercise full and partial matches.
\item[\bmLookupFail:] In a single burst with \PROBESIZE random parameters we test negative lookups (neither key nor value present). We assume this test equivalent to \bmDelete with no match. 
\end{description}

\subsection{Experiment Results}

Figures~\ref{fig:all-benchmarks-clojure-multimap-boxplot} and \ref{fig:all-benchmarks-scala-multimap-boxplot} show the relative differences of specialized \HHAMT multi-map compared to the implementations in Clojure and Scala. From the data we can evaluate our hypotheses:

\begin{description}
\item [Confirmation of Hypothesis 5:] Runtimes unexpectedly improve over the competition. \bmLookup, \bmInsert, and \bmDelete perform similar to Scala (by a median \SI{12}{\percent}, \SI{9}{\percent}, and \SI{16}{\percent} faster), and clearly better than Clojure (by a median speedup of \SI{2.51}{\factor}, \SI{1.75}{\factor}, and \SI{2.05}{\factor}). Compared to Scala we observed individual data points that exhibited minimal slowdowns of less than \SI{9}{\percent} at larger input sizes.

\item [Confirmation of Hypothesis 6:] \HHAMT performs worse than Scala for negative lookups. Runtimes increased by a median \SI{39}{\percent} and roughly doubled at maximum with a \SI{106}{\percent} increase. In contrast, when compared to Clojure we do not see a negative impact.

\item [Confirmation of Hypothesis 7:] Memory footprints improve by a median factor of \SI{1.92}{\factor} (32-bit) and \SI{1.93}{\factor} (64-bit) over the implementation in Scala, and over in Clojure by a median factor of \SI{1.9}{\factor} (32-bit) and \SI{2.14}{\factor} (64-bit).
\end{description} 

\paragraph{Discussion.} We were surprised that the memory footprint consumptions of Clojure's and Scala's multi-map implementations are essentially equal. From related work~\cite{Steindorfer:2015iz} we knew the typical trade-offs of both libraries: Scala mainly optimizes for runtime performance, while Clojure optimizes for memory consumption. Code inspection revealed the cause of Scala's improved memory performance: their immutable hash-sets contains a specialization for singleton sets.

All three libraries libraries follow different paradigms for avoiding code duplication in their collection libraries. While Scala and Clojure use extension mechanisms (i.e., traits and protocols respectively), \HHAMT avoids duplication by supporting internal heterogeneity.

\paragraph{Summary.}
With microbenchmarks we were able to measure the performance of individual operation, and further to measure the footprint of synthetically generated structures of different sizes. In this setting the heterogeneous design of specialized \HHAMT proved to be better in general: improved runtimes of lookup, insertion, and deletion ---with the notable exception of negative lookups when compared to Scala--- and most importantly memory improvements of \SIrange{1.9}{2.14}{\factor}.

\section{Case Study: Primitive Collection Footprint\label{sec:evaluation-primitive-collections}} % \TODO{make plural}

A type-safe heterogeneous \HAMT encoding shines most with bounded numerical data: it allows to exploit the difference between primitive (value-based) data types and reference types. More specifically, a \lstinline{Multimap<int, int>} can leverage storing unboxed inlined singletons. Any non-heterogeneous immutable collection structure would have to store boxed integer objects instead, if not singleton sets of boxed integers. So, instead, as a fair point-of-reference we will compare to the state-of-the-art hand-optimized specialized immutable data structures for primitive values.

We are not aware of any comparable persistent or immutable primitive collection library which is optimized for primitive data types on the \JVM. While there are many specialized primitive collection libraries for the \JVM, only some contain (slower) copy-on-write immutable data structures implemented as facades over their mutable counterparts. With respect to primitive multi-maps, we did not find any implementation, neither mutable nor mutable. 

So, we concentrate on comparing the memory footprint of \lstinline{Map<int, int>}, implemented in specialized \HHAMT (with $1 : 1$ mappings) compared to the most efficient primitive \emph{mutable} collections we are aware of, namely: Goldman Sachs Collections, FastUtil, Trove, and Mahout. As a point of reference we also include Guava's \lstinline{RegularImmutableMap} because it is a well-known library (but commonly known to be non-optimal in terms of memory consumption). 

\subsection{Experiment Results}

\begin{figure}[t]
	\begin{center}
		\includegraphics[width=\columnwidth]{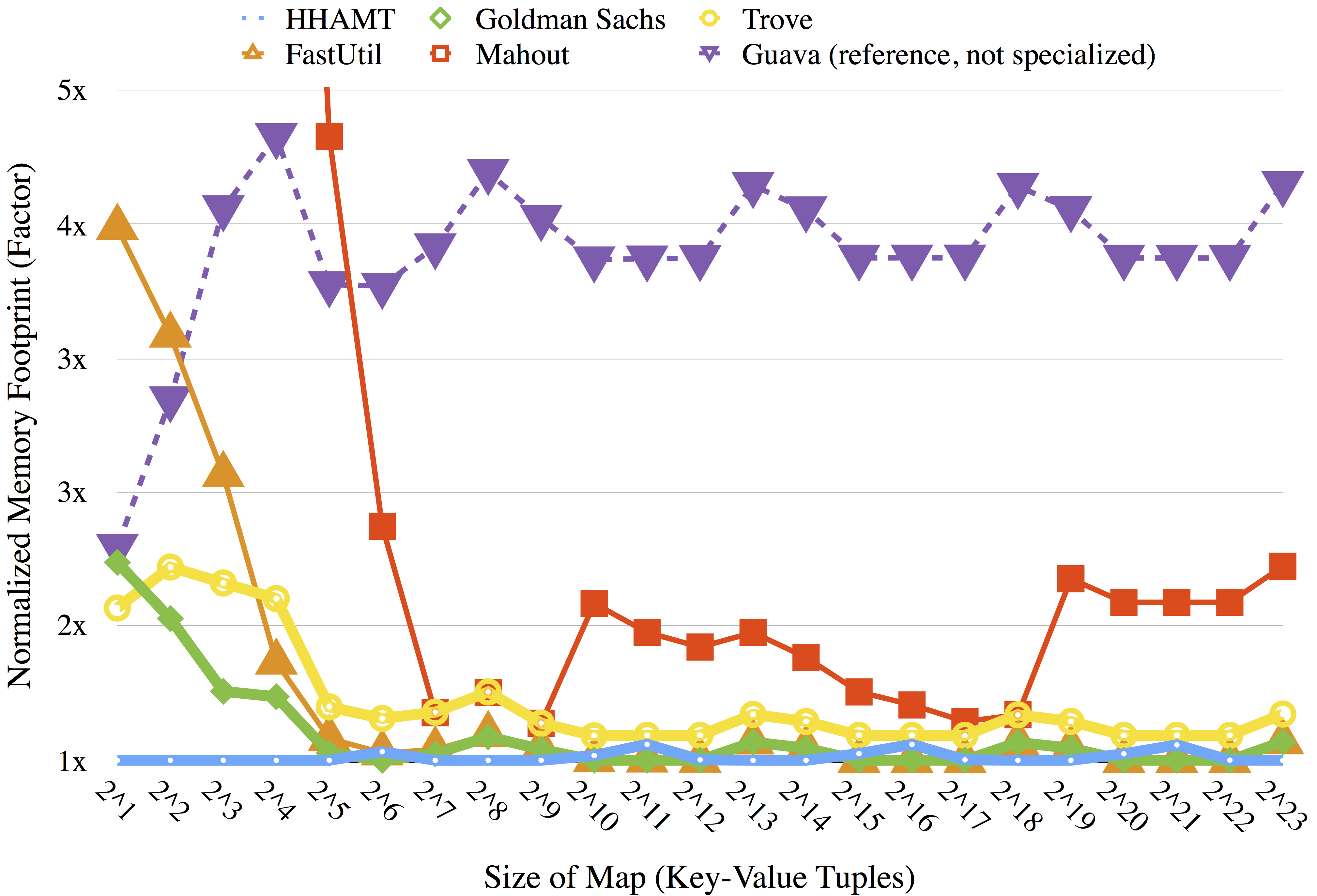}
		\caption{Comparing memory footprint of \HHAMT specialized for \lstinline{int} against state-of-the-art primitive \lstinline{Map<int, int>} structures.}\label{fig:memory_footprint_hhamt_int_int_vs_others}
	\end{center}
\end{figure}

Figure~\ref{fig:memory_footprint_hhamt_int_int_vs_others} illustrates observed memory footprints for maps for sizes $2^x$ for $x \in [1,23]$. At each size, measurements are normalized with respect to the minimum memory footprint (retained size of heap graph). Consequently, the minimum value $1$ depicts the smallest data structure, whereas all other data points are displayed in their relative distance (factor of how much more memory they consume). 

The results show that \HHAMT consistently consumes the least amount of memory (median \SI{1.00}{\factor}, range \SIrange{1.00}{1.10}{\factor}), followed by Goldman Sachs (median \SI{1.04}{\factor}, range \SIrange{1.00}{2.18}{\factor}) and FastUtil (median \SI{1.07}{\factor}, range \SIrange{1.00}{4.18}{\factor}). Trove exhibits constant results within a small bandwidth (median \SI{1.23}{\factor}, range \SIrange{1.15}{2.15}{\factor}). In contrast to Trove's constant results, Mahout delivered surprisingly inconsistent results (median \SI{1.94}{\factor}, range \SIrange{1.22}{29.64}{\factor}) ---we capped the plot in Figure~\ref{fig:memory_footprint_hhamt_int_int_vs_others} to display maximal deviations up to \SI{5}{\factor}. With overheads of \SIlist{29.64;25.08;19.18;11.24;4.72}{\factor} for the data points $2^1$--$2^5$, Mahout exceeds the footprints of our generic reference data structure from Guava (median \SI{4.00}{\factor}, range \SIrange{2.27}{4.72}{\factor}).

\paragraph{Discussion.}
Compared to all other primitive collections, \HHAMT excelled especially at small collections up to \num{32} elements. Given that in practice most collections are small~\cite{Mitchell:2007ex} these improvements look promising. Primitive collections in general have the problem how to mark which slots are in use (there is no \lstinline{null} equivalent in value types). Several encodings ---e.g., sentinel values, or bitmaps--- exist to circumvent this limitation. \HHAMT performs well with respect to primitive collections, because \HHAMT inherently encodes information about the presence and type of (primitive) values on a per node basis an therefore obsoletes special encodings for sentinel values. Further applications and benefits of heterogeneous data structures are discussed in Section~\ref{sec:further-applications}.

\paragraph{Summary.} In our measurements, \HHAMT multi-maps that are specialized for \lstinline{int} consume (with $1 : 1$ data) a median \SI{4}{\factor} less memory than generic map data structures. \HHAMT further achieves the same small footprints as class-leading primitive \lstinline{Map<int, int>} data structures, while providing the additional functionality of allowing $1 : n$ mappings.

% !TEX encoding = UTF-8 Unicode
% !TEX root = ../paper.tex

\section{Case Study: Static Program Analysis\label{sec:evaluation-real-world}}

The above experiments isolate important factors, but they do not show the support for the expected improvements on an algorithm ``in the wild''. To add this perspective, we selected computing control flow dominators using fixed point computation over sets~\cite{aho86}.  The nodes in the graphs are complex recursive ASTs with arbitrarily complex (but linear) complexity for \lstinline{hashCode} and \lstinline{equals}.  More importantly, the effect of the heterogenous encoding does depend on the accidental shape of the data, as it is initially produced from the raw control flow graphs, and as it is dynamically generated by the incremental progression of the algorithm.

\paragraph{Code.}
Although we do not claim the algorithm in this section to be representative of all applications of multi-maps, it is a basic implementation of a well known and fundamental algorithm in program analysis. It has been used before to evaluate the efficiency of hash-array mapped tried~\cite{Steindorfer:2015iz}. We implemented the following  two equations directly on top of the multi-maps:
\begin{align*}
\operatorname{Dom}(n_0) & = \{n_0\} \\
\operatorname{Dom}(n) & = \left(\bigcap_{p \in \operatorname{preds(n)}} \operatorname{Dom}(p)\right) \cup \{n\}
\end{align*}
Our code uses set union and intersection in a fixed-point loop: $\operatorname{Dom}$ and $\operatorname{preds}$ are implemented as multi-maps. The big intersection is not implemented directly, but staged by first producing a set of sets for the predecessors and intersecting the respective sets with each other.

\paragraph{Hypotheses.} On the one hand, since $\operatorname{Dom}$ is expected to be many-to-many with large value sets it should not generate any space savings but at least it should not degenerate the runtime performance either compared to \CHAMP (Hypothesis 8). On the other hand we expect $\operatorname{preds}$ to be mostly one-to-one and we should get good benefit from the inlining of singletons (Hypothesis 9). Since \CHAMP was reported to outperform existing state-of-the-art implementations in Scala and Clojure on the same case, there is no need to further include these~\cite{Steindorfer:2015iz}.

\paragraph{Data.}
For our experiment, we used $\pm$\num{5000} control flow graphs for all units of code (function, method and script) of Wordpress,\footnote{\url{https://wordpress.com}} by using the PHP AiR framework~\cite{phpair}.
Like before, we used \JMH to measure \CPU time. We ran the dominator calculations on a random selection of the aforementioned graphs. The set of selected graphs range between a size of from \num{128} to \num{4096} in exponential steps. Since smaller graphs occur much more frequently, we selected samples with exponentially increasing sizes from \num{128} to \num{4096}. We furthermore measured the number of many-to-one and many-to-many entries in the $\operatorname{preds}$ relation.

\newcommand{\realWorldBenchmarkTable}[3]{%
		\caption{#1}%
		\vspace{1mm}
    	\small%\footnotesize%
    	\ra{1.4}%
    	\begin{tabularx}{\linewidth}{@{\hspace*{2pt}}r*{4}{r}*{1}{R}}%
    		\toprule%
%				\multirow{2}{*}{\tableheader{\#\CFG}} & \multirow{2}{*}{\tableheader{Clojure}} & \multirow{2}{*}{\tableheader{Scala}} & \multirow{2}{*}{\tableheader{\HHAMT}} & \multicolumn{2}{c}{\tableheader{Speedup w.r.t.} & \multicolumn{2}{c}{\tableheader{Speedup w.r.t.}} \\%
%				\cmidrule(l){5-6}%
%				\multicolumn{4}{c}{} & \tableheader{Clojure} & \tableheader{Scala} & \tableheader{Scala} \\%
\tableheader{\#\CFG} & \tableheader{\CHAMP} & \tableheader{\HHAMT} & \tableheader{\#Keys} & \tableheader{\#Tuples} & \tableheader{\% 1 : 1} \\ 
    		\midrule%
    		#2%
%    		\midrule%
%			#3%
    		\bottomrule%
    	\end{tabularx}%
}%
%%%
%%%
%%%
%\FloatBarrier
\begin{table}
\realWorldBenchmarkTable{Runtimes of \HHAMT for the \CFG dominators experiment per \CFG count, and statistics over $\operatorname{preds}$ relation about shape of data (unique keys, tuples, $1 : 1$ mappings).}%
	{% !TEX encoding = UTF-8 Unicode
% !TEX root = ../paper.tex

\num{4096} & \SI{173}{\second} & \SI{174}{\second} & \num{315009} & \num{331218} & \SI{91}{\percent} \\ % & \SI{1.05}{\factor} & \num{302371}
\num{2048} & \SI{84}{\second} & \SI{85}{\second} & \num{162418} & \num{170635} & \SI{91}{\percent} \\ % & \SI{1.05}{\factor} & \num{155968}
\num{1024} & \SI{64}{\second} & \SI{62}{\second} & \num{88952} & \num{93232} & \SI{92}{\percent} \\ % & \SI{1.05}{\factor} & \num{85617}
\num{512} & \SI{28}{\second} & \SI{28}{\second} & \num{43666} & \num{45743} & \SI{92}{\percent} \\ % & \SI{1.05}{\factor} & \num{42062}
\num{256} & \SI{19}{\second} & \SI{18}{\second} & \num{21946} & \num{22997} & \SI{92}{\percent} \\ % & \SI{1.05}{\factor} & \num{21140}
\num{128} & \SI{14}{\second} & \SI{14}{\second} & \num{13025} & \num{13583} & \SI{93}{\percent} \\ % & \SI{1.04}{\factor} & \num{12584}
}{}%
	\label{tbl:real-world-chart-vs-clojure}%
\vspace{4mm}
\end{table}
%\FloatBarrier
%%%
%%%
%%%

\paragraph{Results.}

The results were obtained with a Linux machine running Fedora~20 (kernel 3.17). It featured \SI{16}{\giga\byte} RAM and an Intel Core i7-2600 CPU with \SI{3.40}{\giga\hertz} (\SI{8}{\mega\byte} \LLC with \num{64}-byte cache lines). Frequency scaling was disabled.

Table~\ref{tbl:real-world-chart-vs-clojure} shows the mean runtimes of the experiment for \CHAMP and \HHAMT. Both perform almost identically, with at most $\pm$\SI{2}{\second} difference. Due to equal runtimes, \HHAMT retains the same magnitude of speedups that \CHAMP yielded over Clojure and Scala~\cite{Steindorfer:2015iz}, from minimal \SI{9.9}{\factor} to \SI{28.1}{\factor}. We also observed that the shape of data in the $\operatorname{preds}$ relation contains a high number of $1 : 1$ mappings (median \SI{92}{\percent}) and that the average ratio of unique keys to tuples is \SI{1.05}{\factor}. In the case of Wordpress, the \CFG algorithm turns out to profit over \CHAMP in terms of memory savings from the heterogeneous opimizations for $1 : 1$ mappings. We conclude both Hypothesis 8 and 9 to be confirmed.

% !TEX encoding = UTF-8 Unicode
% !TEX root = ../paper.tex

\section{Related Work\label{sec:related_work}}

\paragraph{Reducing the Memory Footprint of Collections} is a goal of other people as well.  
Gil et al.~\cite{Gil:2012gq} identified sources of memory inefficiencies in Java's mutable collections and proposed memory compaction techniques to counter them. They improved the memory efficiency of Java's \texttt{Hash\{Map,Set\}} and \texttt{Tree\{Map,Set\}} data structures by \SIrange{20}{77}{\percent}. We observed that even with added heterogeneity, \HHAMT multi-maps achieve lower memory footprints than the class-leading primitive collection libraries, and in the generic case on average \SI{4}{\factor} improvements over Guava's maps.

Steindorfer and Vinju~\cite{Steindorfer:2014dk} specialized internal trie nodes to gain memory savings of \SI{55}{\percent} for maps and \SI{78}{\percent} for sets at median while adding \SIrange{20}{40}{\percent} runtime overhead for lookup. Their approach minimized the amount of specializations to mitigate effects on code bloat and run-time performance. In contrast, we targeted the root causes of inefficiency one-by-one allowing full specialization at all arities.

\paragraph{Optimizing Collections in Dynamically-Typed Languages.} Runtimes of dynamically typed languages often introduce a significant run-time and memory overhead~\cite{Tratt:2009bn} due to generic collection data structures that could at run-time hold a heterogeneous mix of data.
\sloppypar Bolz et al.~\cite{Bolz:2013ku} introduced a technique dubbed \emph{storage strategies} that enables dynamic conversion of data representations. A set of interlinked strategies form a fine-grained type lattice that is based on known optimizations. Strategies mostly include support for collections of a homogeneous (primitive) type. An exemplary lattices for a \lstinline{Set} data structure could be \lstinline{EmptySetStrategy <-> (Integer|Float|...)SetStrategy <-> ObjectSetStrategy}. Resulting performance improvements mainly stem from object layouts that specialize for a homogeneous primitive types and corresponding optimized operations (e.g., direct value comparisons instead of calling \texttt{equals} methods).

Bolz~\cite{Bolz:2013ku} showed that with Python on average 10\% of collections dehomogenize, mostly at small sizes. These results suggest that even in the absence of strict typing, collections are often used homogeneously. 
Heterogeneous data structures are orthogonal to homogeneous storage strategies. On one hand, heterogeneous data structures could diversify current strategy approaches, e.g., when homogeneous strategies are not applicable, or when many conversion occur. On the other hand, they have the potential to replace homogeneous strategies when flexibility in mixing data is required upfront.
Furthermore, \HHAMT optimizes internal heterogeneity that occurs in general purpose data structures such as multi-maps.

\paragraph{Specializations and Generics for Primitives Reducing Code-Bloat.} 
Specializing for primitives can lead to a combinatorial explosion of variants amplifying code-bloat. Due to the object vs. primitive dichotomy, Java  does not offer solutions countering a combinatorial explosion of code duplication when specializing for primitives. Java~10 or later will solve this issue by supporting generics of primitive types.\footnote{\url{http://openjdk.java.net/projects/valhalla/}}

Ureche et al.~\cite{Ureche:2013gj} presented a compiler-based specialization transformation technique called \emph{miniboxing}. Miniboxing adds automatic specializations for primitive \JVM data types to the Scala compiler while reducing the generated bytecode. Combinatorial code-bloat is tackled by specializing for the largest primitive type \lstinline{long}, together with automatic coercion for smaller-sized primitives. While not always memory-optimal due to always utilizing \lstinline{long} variables, miniboxing is a practical approach to tackle combinatorial code explosion. 

\HHAMT's contribution is orthogonal to both previously discussed techniques, because it generalizes the encoding of heterogeneous data stored together in a collection. \HHAMT's specializations currently do duplicate code for different (primitive) type combinations. Using primitive generics in later versions of Java ---or miniboxing in Scala--- could bring this down to a single generic specialization per trie node arity.

\paragraph{(Partial) Escape Analysis.} Escape analysis enables compilers to improve the run-time performance of programs: it determines whether an object is accessible outside its allocating method or thread. Subsequently this information is used to apply optimizations such as stack allocation (in contrast to heap allocation), scalar replacements, lock elision, or region-based memory management~\cite{Stancu:2015:SEH:2754169.2754185}. Current \acp{JVM} use partial escape analysis~\cite{partical-escape-analysis-stadler-thesis}, which is a control-flow sensitive and practical variant tailored toward \JIT compilers.

Our scalable encoding of specializing is a memory layout optimization for value-type based data types: trie nodes are specialized for arrays of constant sizes that do not escape. We use code generation to conceptually apply object inlining~\cite{object-inlining} of these statically sized (non-escaping) arrays into the memory layout of their corresponding trie nodes. Memory layout sensitive inlining as we perform could be applied in \VM based on information obtained from escape analysis. We hope that future compilers and language runtimes are capable of doing so out-of-the-box.
% !TEX encoding = UTF-8 Unicode
% !TEX root = ../paper.tex

\section{Further Applications of \HHAMT\label{sec:further-applications}}

We extrapolate some client applications which would benefit from \HHAMT.

\paragraph{Libraries or Languages Supporting Data or Code Analysis on the \JVM} would benefit from more efficient in-memory multi-maps. Typical examples are frameworks such as KNIME~\cite{Berthold:2009:KKI:1656274.1656280} for general purpose data analysis or Rascal for program analysis~\cite{rascal}, and MoDisCo~\cite{Bruneliere:2010:MGE:1858996.1859032} for software re-engineering and reverse engineering, especially when their algorithms require frequent lookup and thus will benefit from an efficiently indexed relation such as a multi-map.

\paragraph{Unifying Primitive and Generic Collections.} Looking at specialized collections for primitives from the programming language designer's perspective, they are a necessary evil implied by the dichotomy between objects and primitive values. Primitive values give programmers access to low level and memory-efficient data representations, but the impact of having them leaks through in the type systems and the design of standard libraries of programming languages supporting them. The current paper describes a heterogeneous framework that can be used for implementing data structure which allow storing either primitive data values or their boxed counterparts next to each other, while the client code remains practically oblivious. For statically-typed languages this implies we can have a generically typed library for both primitive and object values. For dynamically-typed languages it implies a much lower overhead for the builtin dictionaries. 

\paragraph{Big Integers for Big Data.} Most programming languages feature a library for representing arbitrary-sized integers. We use these to avoid overflow, especially in the context of scientific computing applications. The drawback of using these libraries for data science is that large homogeneous collections immediately blow up, even if the big numbers are exceptional. We want to use smaller \acp{FIXNUM} were possible, and \acp{BIGNUM} only when necessary.

This application is where \HHAMT could potentially have a rather big impact. Sets and maps filled with mostly inlined \FIXNUM's and an occasional \BIGNUM without having to a priori allocate space for \acp{BIGNUM}, and without having to migrate at run-time. Even if the entire collection accidentally ends up filled with \acp{BIGNUM}, \HHAMT still yields more memory efficient representations than common array-based hash-maps.

\paragraph{Cleaning Raw Data in a Type-Safe Manner.} The \HHAMT data structure enables efficient storage and retrieval of objects of incomparable types without memory overhead (no need to wrap the objects) and without dynamic type checks. In Java there exist no ``union'' types like in C, but using \HHAMT we can approach this in the context of collections. A typical use case would be reading in raw data from \CSV files (or spreadsheets) in Java where the data is not cleansed and some cells contain integers while the other contain decimal numbers or even empty cells, depending on the original manual and unvalidated input of the user. A \CSV parser could output a \HHAMT, inferring the most accurate value for each cell from the used notation, and allowing for further processing the data downstream in a manner both type-safe and efficient.

In general, homogeneous collections storing numeric data struggle with representing empty cells. Sentinel values (e.g., integer constant \num{-1}) are a viable solution if and only if the data does not use the data type's full value range. Cases where the data range is used exhaustively require additional auxiliary data structure (e.g., an array of booleans) to encode if a value is initialized. In contrast to homogeneous collections, \acp{HHAMT} by design supports mixing  sentinel values of a different type (e.g., \lstinline{static final EMPTY_CELL = new Object()}) with the full value range of primitives.
% !TEX encoding = UTF-8 Unicode
% !TEX root = ../paper.tex

\section{Conclusion\label{sec:conclusion}}

We proposed \HHAMT, a new design for hash-array mapped tries which allows storage, retrieval and iteration over maps which store heterogeneously typed data. In particular we motivate this data structure by applying it to efficiently implement multi-maps, and it also shines when used to cater for inlining unboxed primitive values. 

The evaluation compared to the state-of-the-art: comparing to other hash-trie data structures with and without the many-to-many feature, comparing against state-of-the-art encodings of multi-maps in Scala and Clojure and comparing to hand-optimized maps for primitive values. Even when compared unfairly to implementations which do not feature heterogeneity, \HHAMT compares well. We safe a lot of memory (\SIrange{2}{4}{\factor}) at relatively low costs in runtime overhead. 

We hope multi-maps based on these results will be available in the future in the standard libraries for collections on the \JVM, since that would make the \JVM even more attractive for safely computing with large immutable datasets.

%\acks
%We thank \dots and our colleagues and the anonymous referees for providing feedback on earlier drafts of this paper.

%\clearpage
%\pagebreak

\bibliographystyle{abbrvnat}
\bibliography{paper}

\appendix
%\clearpage
% !TEX encoding = UTF-8 Unicode
% !TEX root = ../paper.tex

\begin{figure*}[t]
\begin{lstlisting}[keepspaces=true, escapechar=!,label=lst:generalized-hhamt-interface]
interface HeterogeneousMap { 
  // pull-based dispatch on type !\label{line:pull}!
  <K, V> TypedObject<?> put    (Class<K> keyType, K key, Class<V> valueType, V value);
  <K, V> TypedObject<?> remove (Class<K> keyType, K key);
  <K, V> TypedObject<?> get    (Class<K> keyType, K key);   
  
  // push-based dispatch on type
  <K, V> void put    (Class<K> keyType, K key, Class<V> valueType, V value, CallbackMap callbacks);
  <K, V> void remove (Class<K> keyType, K key, Class<V> valueType, V value, CallbackMap callbacks);
  <K, V> void get    (Class<K> keyType, K key, Class<V> valueType, V value, CallbackMap callbacks);
}

interface TypedObject<T> {
  Class<T> getType();
  T get();
}

interface CallbackMap {
  <E> Consumer<E> put (Class<E> elementType, Consumer<E> consumer);
  <E> Consumer<E> get (Class<E> elementType);
}	
\end{lstlisting}
	\caption{Generic \HHAMT interface, based on \emph{Item 29: Consider typesafe heterogeneous containers} of \emph{Effective Java}~\cite{effective-java}. \label{lst:api}}
\end{figure*}

\begin{figure*}
\begin{lstlisting}[keepspaces=true, label=lst:generalized-hhamt-interface-usage]
public void heterogeneousInterfaceTest() {
  put(String.class, "abc", int.class, 5);      // accepted by guard condition
  put(String.class, "abc", Integer.class, 5);  // accepted by guard condition
  
  put(String.class, "abc", long.class, 5L);    // rejected by guard condition
  put(String.class, "abc", Long.class, 5L);    // rejected by guard condition
}  
  
static <T, U> void put(Class<T> keyType, T keyInstance, Class<U> valueType, U valueInstance) {  
  switch(keyType.getName()) {
    case "java.lang.String":
      switch(valueType.getName()) {
        case "int":
          put((String) keyType.cast(keyInstance), (int) valueInstance);
          return;
        case "java.lang.Integer":
          put((String) keyType.cast(keyInstance), (Integer) valueInstance);
          return;
      }
  }
        
  System.out.println("Unsupported Type");
}

static void put(String keyInstance, Integer valueInstance) {  
  System.out.println("put(String keyInstance, Integer valueInstance)");
}

static void put(String keyInstance, int valueInstance) {  
  System.out.println("put(String keyInstance, int valueInstance)");
}
\end{lstlisting}
	\caption{The method \lstinline{heterogeneousInterfaceTest} illustrates a possible way to map a generalized \HHAMT interface to specialized functions with type guards (cf.~\lstinline{switch} statement).}
\end{figure*}

%\clearpage
%\input{chapters/heterogeneous_support_for_champ.tex}
%\clearpage
%\input{chapters/cover_letter.tex}

%% The bibliography should be embedded for final submission.
%
%\begin{thebibliography}{}
%\softraggedright
%
%\bibitem[Smith et~al.(2009)Smith, Jones]{smith02}
%P. Q. Smith, and X. Y. Jones. ...reference text...
%
%\end{thebibliography}

\end{document}